\documentclass[a4paper,11pt]{article}
\pdfoutput=1 

\usepackage{jheppub}
\usepackage{color} 
\usepackage{float}
\usepackage{ulem}
\usepackage{braket}
\usepackage{amsmath,graphicx,amssymb,subfigure,bbm}
\usepackage[all]{xy}

\begin{document}

\title{Vector mesons from a holographic QCD model in $f(R)$-dilaton gravity}

\author[a]{Ad\~ao S. da Silva Junior,}
\author[b]{Juan M. Z. Pretel}

\affiliation[a]{Instituto de F\'{i}sica, Universidade
Federal do Rio de Janeiro, \\
Caixa Postal 68528, RJ 21941-972, Brazil.}

\affiliation[b]{Centro Brasileiro de Pesquisas F{\'i}sicas, Rua Dr.~Xavier Sigaud, \\
150 URCA, Rio de Janeiro CEP 22290-180, RJ, Brazil}
\emailAdd{adaofisic@gmail.com}
\emailAdd{juan04manuel91@gmail.com }

\abstract{In this paper we investigate the spectrum and decay constants of vector mesons based on $f(R)$-dilaton gravity. We focus particularly on the well-known Starobinsky model, given by the function $f(R)= R+ \alpha R^2$ in the metric formalism, and we examine the deviations from the pure Einstein-dilaton case. We thus provide the field equations for Starobinsky-dilaton gravity in $\rm AdS_{5}$ spacetime and calculate the spectral decomposition of the vector mesons by means of a Sturm-Liouville problem. Remarkably, for the rho mesons $m_{\rho^2}$ and $m_{\rho^3}$, our results are in good agreement with the experimental data when $\alpha= 10^{-8}$. Our work also generalizes previous studies and recovers standard results when $\alpha= 0$. }

\maketitle

\flushbottom

\section{Introduction}
The AdS/CFT correspondence is a duality between string theories on $\text{AdS}_{d + 1}\times \mathcal{M}$ manifold and conformal field theory \cite{Maldacena:1997re, Witten:1998qj, Gubser:1998bc} (see also the excellent book \cite{Ammon:2015wua}). In the present study we will focus on the equivalence between $\mathcal{N} = 4$ super Yang Mills theory in $3 + 1$ dimensions and IIB string theory on $\text{AdS}_{5}\times S^{5}$ spacetime. This conjecture establishes a mapping between a classical gravitational theory in a weakly curved spacetime (the AdS side of the conjecture) and a strongly coupled field theory (the CFT side). This property has enabled significant progress in studying the non-perturbative regime of Quantum Chromodynamics (QCD), which, although well understood in the perturbative regime, poses significant theoretical challenges in the hadronic regime.

Based on the gauge/gravity duality, bottom-up AdS/QCD models (such as hard-wall and soft-wall) have been widely used to describe various properties of hadrons, such as mass spectra, decay constants, form factors, radii, and coupling constants \cite{Karch:2006pv, Abidin:2008hn, Abidin:2008ku, Abidin:2009aj, DaRold:2005mxj, Grigoryan:2007my, Grigoryan:2007vg, Grigoryan:2007wn}. Within these approaches, there is a class known as Einstein-dilaton models. This model in its zero-temperature version provides a dynamics for the dilaton field, has a nontrivial vacuum, and establishes a general confinement criterion \cite{Gursoy:2007cb, Gursoy:2007er}. In its finite-temperature version, it very successfully describes the confinement-deconfinement transition and thermodynamics, satisfying the confinement properties \cite{Gursoy:2008za}. This model has been applied in several contexts, describing mass spectra and trace anomaly, chiral symmetry breaking, calculation of shear viscosity, bulk viscosity and drag force \cite{Ballon-Bayona:2017sxa, Ballon-Bayona:2023zal, Ballon-Bayona:2021tzw, Gursoy:2009kk}.

Higher curvature theories, such as Lovelock, Gauss-Bonnet and Horndeski, have been widely applied in holography to explore various phenomena, including entanglement entropy, violations of the viscosity bound, thermal phase transitions and thermodynamics of BTZ black holes within the framework of AdS/BCFT \cite{Hung:2011xb, Brigante:2007nu, Cho:2002hq, Santos:2021orr}. Of course, the literature offers a wide variety of modified gravity theories, but the simplest way to modify Einstein's General Relativity is by replacing the Ricci scalar $R$ in the standard Einstein-Hilbert action by a generic function of $R$, namely, the well-known $f(R)$ gravity theories \cite{Sotiriou2010, DeFelice2010}. From a cosmological point of view, $f(R)$ gravity has been formulated as a geometric alternative to explain phenomena such as inflation and late-time acceleration, see e.g.~Refs.~\cite{Capozziello2011, Nojiri2011, Clifton2012, Nojiri2017, NOJIRI2020} for a comprehensive review. In fact, by choosing the appropriate function $f(R)$, the accelerated expansion of the Universe arises naturally without the need to introduce an exotic fluid component like dark energy \cite{Carroll2004, Song2007, Ali2010, Ribeiro2024}. This and other motivations recommend us to go beyond conventional Einstein gravity so that the Einstein-dilaton model has to be extended to include higher-order curvature terms in its action.

The simplest function in $f(R)$ gravity is given by $f(R)= R+ \alpha R^2$ (with $\alpha >0$ in order to be free of tachyonic instabilities), the first model of inflation proposed by Starobinsky in 1980 \cite{Starobinsky1980}. Such a model is also known as $R$-squared gravity and provides a good fit to the \textit{Planck} 2018 data via the statistical properties of the cosmic microwave background (CMB) anisotropy measurements \cite{Akrami2020}. On the other hand, from the context of relativistic compact stars, this gravity model has been confronted with different astrophysical measurements providing promising results \cite{Astashenok2017, Olmo2020, Astashenok2020, Astashenok2021, Nobleson2022, Jimenez2022, Pretel2022JCAP}. Black hole solutions in $R^2$-gravity have also been reported in the literature \cite{Kehagias2015}, as well as the thermodynamics of topological black holes in that gravitational context \cite{Cognola2015}. Furthermore, it is remarkable to mention that, based on the AdS/CFT correspondence, holographic images of an AdS black hole under $f(R)$ gravity have been recently obtained by using wave optics \cite{Li2024}. Motivated by these investigations, in this work we aim to discuss the impact of the Starobinsky gravity model on the $\rho$ vector meson spectrum and decay constants.  Against this background, we show that it is possible to include quadratic corrections $\alpha R^2$ to describe a spectrum compatible with the experimental data. In addition, we calculate the decay constants of the vector mesons. Our results indicate that the effects of Starobinsky gravity become more pronounced for higher excited states. To the best of our knowledge, this work constitutes the first study of the vector meson spectrum and decay constants within the framework of Starobinsky-dilaton gravity.

As will be shown in our results, the spectrum becomes nonlinear because the more highly excited states are more strongly affected by the $\alpha$-corrections. The non-linearity of the meson spectrum has been explored in the context of soft-wall models \cite{MartinContreras:2020cyg, Braga:2025lah, Afonin:2009xi}, and also in a proposed independent model \cite{Chen:2022flh}. Our goal in extending Einstein-dilaton gravity to the Starobinsky-dilaton framework is to improve the phenomenological description of the vector meson spectrum within a gravitationally consistent setup. In this regard, the $\alpha R^{2}$ correction introduces deviations from linear Regge behavior that emerge dynamically from the geometry, rather than being imposed by construction. As we will see later, these corrections predominantly affect the highly excited states, leading to a nonlinear spectrum. We emphasize that our approach does not aim to strictly enforce asymptotic Regge linearity, but rather to explore the impact of higher-curvature corrections on the spectrum while maintaining an essential physical property, namely the confinement criterion. In contrast, soft-wall models such as in Ref.~\cite{Karch:2006pv} achieve linearity by construction through a quadratic infrared behavior of the effective Schrödinger potential, but do not satisfy the confinement criterion associated with the Wilson loop.

The structure of this paper is as follows. In Section \ref{section2}, we construct the model by introducing an $f(R)$ modification into the original Einstein-dilaton action in a five-dimensional (5D) anti-de Sitter space. We derive the field equations using the metric formalism and adopt Starobinsky gravity for the modified gravitational field equations. At the end of this section, we examine the impact of quadratic corrections through the free parameter $\alpha$ on confinement. In Section \ref{section3}, we present the action and field equations for vector mesons. Subsequently, we analyze the potential, mass spectrum, wave functions and decay constant, incorporating the effects of the Starobinsky model correction for some selected values of $\alpha$. The rho meson spectrum is compared with the experimental data. Finally, the main conclusions of this paper are presented in Section \ref{section4}.

\section{Holographic QCD model from $f(R)$-dilaton gravity}\label{section2}

\subsection{The action and field equations}
Motivated by metric $f(R)$ gravity theories \cite{Sotiriou2010, DeFelice2010}, we consider the modified Einstein-dilaton action in the Einstein frame written as
\begin{align}
    S_{E} = \sigma\, \int d^{5}x \sqrt{g}\left[f(R) - \dfrac{4}{3}\partial^{m}\Phi\partial_{m}\Phi + \ell^{-2}V(\Phi)\right]\,, \label{action}
\end{align}
where $\sigma = 1/16 G_{5}$ with $G_{5}$ being the five dimensional Newton's constant, $f(R)$ is a generic function of the Ricci scalar $R$, $\Phi$ is the dilaton field, $V(\Phi)$ is the dilaton potential and $\ell$ is the AdS$_{5}$ radius. Under the metric formalism, the field equations are obtained by varying such action with respect to the metric tensor and dilaton field. Thus, the field equations in this formalism are given by
\begin{align}
    f_{R}R_{mn} - \dfrac{1}{2}g_{mn}f(R) + \left(g_{mn}\nabla^{2} - \nabla_{m}\nabla_{n}\right)f_{R} &= \dfrac{4}{3}\partial_{m}\Phi\partial_{n}\Phi + \dfrac{1}{2}\mathcal{L}_{\Phi}g_{mn}\label{eqmotionfr}\,,\\
    \nabla^{2}\Phi + \dfrac{3}{8\ell^{2}}\dfrac{dV}{d\Phi} &= 0\, ,\label{dilatoneq}
\end{align}
where $f_{R}(R)= df(R)/dR$ and $\nabla^{2}$ is the is the d'Alembert operator. The dilaton equation \eqref{dilatoneq} can also be obtained using the Bianchi identity $\nabla_{n}T^{mn} = 0$. The trace of the above equation is obtained by contracting it with $g^{mn}$, namely
\begin{equation}
    4 \nabla^{2}f_{R} + R f_{R} - \dfrac{5}{2}f(R) = - 2\partial^{n}\Phi\partial_{n}\Phi + \dfrac{5}{2\ell^{2}}V(\Phi)\,, \label{traceeq}
\end{equation}
which shows the dynamics of the Ricci scalar through the $f(R)$ contribution in the presence of the dilaton field $\Phi$. Particularly, when $f(R) = R$ (i.e., $f_{R} = 1$), equations \eqref{action}, \eqref{eqmotionfr} and \eqref{traceeq} are reduced to the conventional Einstein-dilaton equations.

Let us consider as background the $5d$ metric
\begin{equation}\label{5Dmetric}
    ds^{2} = e^{2A(z)}[dz^{2} + dx_{i}^{2} - dt^{2}]\,,
\end{equation}
where $A(z)$ is the warp factor and $i = 1, 2, 3$. Substituting this metric into the $f(R)$-dilaton equation \eqref{eqmotionfr}, we find the following set of equations:
 \begin{align}
   R^{\prime\prime}f_{RR} + R^{\prime\,\, 2}f_{RRR} + 6 A^{\prime}R^{\prime}f_{RR} - e^{2A}(f + \ell^{-2}V(\Phi) - R f_{R}) +  3(A^{\prime\prime} + 3A^{\prime\, 2})f_{R} = 0\,, \label{fReqmotion1}\\
   4\left(R^{\prime\, \prime}f_{RR} + R^{\prime\, 2}f_{RRR} + \Phi^{\prime\, 2}\right) + 12 A^{\prime\prime}f_{R} - e^{2A}(f + \ell^{-2}V(\Phi) - Rf_{R}) = 0\,, \label{fReqmotion2}\\
   4\left(R^{\prime\, \prime}f_{RR} + R^{\prime\, 2}f_{RRR} + 3 A^{\prime}R^{\prime}f_{RR}\right) + e^{2A}R f_{R} - \dfrac{5}{2}e^{2A}(f + \ell^{-2}V(\Phi)) + 2\Phi^{\prime\, 2} = 0\,, \label{fReqmotion3}\\
   \dfrac{8}{3}\left(\Phi^{\prime\prime} + 3 A^{\prime}\Phi^{\prime}\right) + \dfrac{e^{2A}}{\ell^{2}}\dfrac{dV}{d\Phi} = 0\,,\label{fReqmotion4}
 \end{align}
where $^\prime = d/dz$. It is worth noting that, in the limit $f(R)=R$ (so that $f_{R}=1$), the above system reduces to the standard Einstein-dilaton equations \cite{Gursoy:2007cb}. In the next section, we will discuss the particular $f(R)$ model employed to calculate the masses of vector mesons.

\subsection{Starobinsky-dilaton gravity}

In order to determine the vector meson masses within a $f(R)$-dilaton gravity context, it is necessary to specify the function $f(R)$. For this purpose, we will use here the so-called $R$-squared gravity (also known in the literature as the Starobinsky model \cite{Starobinsky1980}), given by
 \begin{equation}
     f(R) = R + \alpha R^{2}\label{starobinsky}\,,
 \end{equation}
where $\alpha$ is a free parameter of the model. Such a non-linear model is cosmologically well motivated because its inflationary predictions are in good agreement with \textit{Planck} 2018 measurements of the CMB anisotropies \cite{Akrami2020}. Accordingly, plugging \eqref{starobinsky} in Eqs.~\eqref{fReqmotion1}-\eqref{fReqmotion4}, we get
 \begin{align}
     3\left(A^{\prime\, \prime} + 3 A^{\prime\, \, 2}\right)\left(1 + 2\alpha R\right) + 2\alpha(R^{\prime\, \prime} + 6 A^{\prime} R^{\prime}) = e^{2A}(\ell^{-2}V(\Phi) - \alpha R^{2})\,, \label{eqstarobinsky1}\\
     12 A^{\prime\prime}(1 + 2\alpha R) +  4(2\alpha R^{\prime\prime} + \Phi^{\prime\,\, 2}) = e^{2A}(\ell^{-2}V(\Phi) - \alpha R^{2})\,, \label{eqstarobinsky2}\\
     16\alpha(R^{\prime\prime} +  3A^{\prime}R^{\prime}) + 2R(1 + 2\alpha R)e^{2A} -  5(R + \alpha R^{2} + \ell^{-2}V(\Phi))e^{2A} + 4\Phi^{\prime\,\, 2} = 0\,, \\
     \dfrac{8}{3}(\Phi^{\prime\prime} + 3A^{\prime}\Phi^{\prime}) + \dfrac{e^{2A}}{\ell^{2}}\dfrac{dV}{d\Phi} = 0\,.
 \end{align}
From Eqs.~\eqref{eqstarobinsky1} and \eqref{eqstarobinsky2}, we can obtain 
\begin{equation}
     6\alpha (2A^{\prime}R^{\prime} - R^{\prime\prime}) + 9(1 + 2\alpha R)(A^{\prime\, 2} - A^{\prime\prime}) - 4\Phi^{\prime\, 2} =0\,. \label{eqstarobinsky3}
\end{equation}

For the five-dimensional metric \eqref{5Dmetric}, the Ricci scalar can be written as 
\begin{equation}
     R = - 4\, e^{-2A}\left(2A^{\prime\prime} + 3A^{\prime\, 2}\right)\,.\label{ricciscalar}
\end{equation}
It is convenient here to write the warp factor as
\begin{equation}
    \zeta(z)\equiv \exp\left[ - A(z)\right]\,. \label{zeta}
\end{equation}
In view of Eqs.~\eqref{ricciscalar} and \eqref{zeta}, Eq.~\eqref{eqstarobinsky3} becomes
\begin{equation}
     4\Phi^{\prime\, 2}\zeta = 24\alpha \zeta^{\prime\, 2} \zeta^{\prime\prime} + 48\alpha \zeta^{\prime\prime\prime}\zeta^{\prime}\zeta - 48\alpha\zeta^{\prime\prime\prime\prime}\zeta^{2} + 336\alpha \zeta^{\prime\prime\, 2}\zeta + 9\zeta^{\prime\prime}\,, \label{eqstarobinsky4}
\end{equation}
and when $\alpha = 0$ we retrieve the standard equation used in Einstein-dilaton gravity, i.e.,
\begin{equation}
      \zeta^{\prime\prime} - \dfrac{4}{9}\Phi^{\prime\, 2}\zeta = 0\,. \label{edpure}
\end{equation}

\begin{figure}[htp!]%
    \centering
    {{\includegraphics[width=7.4cm]{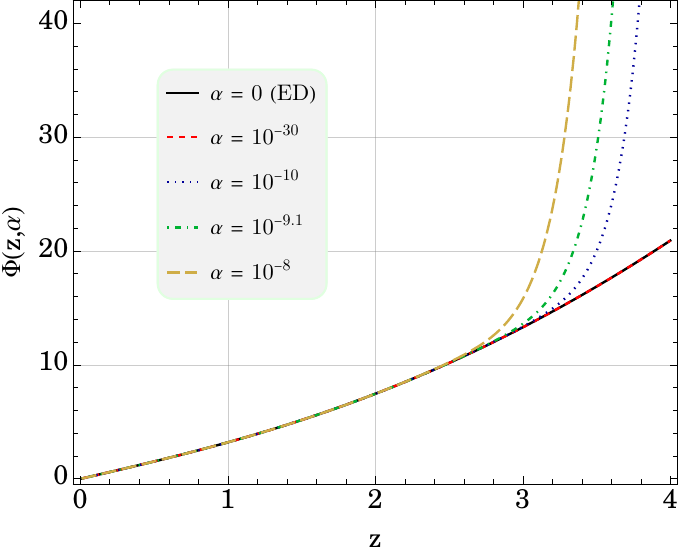}}
    {\includegraphics[width=7.42cm]{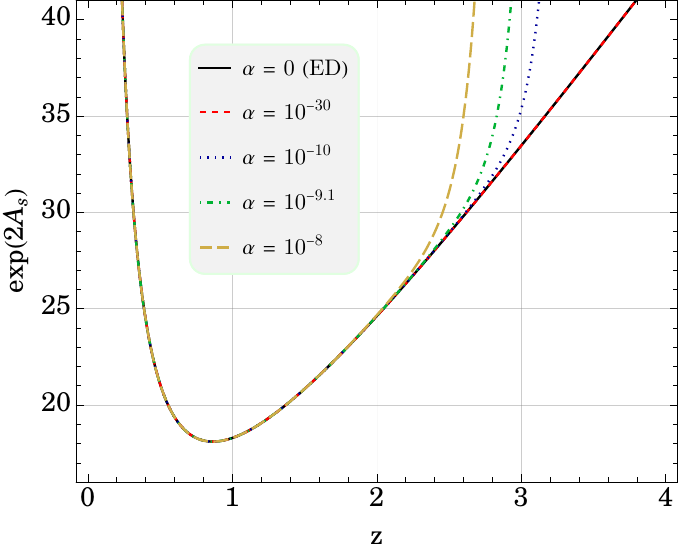}}}%
    \caption{Dilaton profile (left) and warp factor (right) as a function of the radial coordinate $z$ in Starobinsky-dilaton gravity for several values of the free parameter $\alpha$. The particular case $\alpha= 0$ corresponds to Einstein-dilaton gravity and has been included as a benchmark by a black solid line. In both plots we have considered units in which $k = 1$.}%
    \label{Plot:DilatonWarpfactor}%
\end{figure}

The last expression was used in Ref.~\cite{Ballon-Bayona:2017sxa} to study glueball spectrum and trace anomaly in the context of Einstein-dilaton gravity, and its generalized version is now given by \eqref{eqstarobinsky4}. Note that Eq.~\eqref{eqstarobinsky4} is a fourth-order differential equation in the inverse scale factor $\zeta$. To solve such an equation, we will use the following inverse scale factor
\begin{equation}
    \zeta(z) = z\exp\left(\dfrac{2}{3}k z^{2}\right)\,, \label{modelii}
\end{equation}
where $k$ sets the mass scale. When $k = 0$, the AdS solution is recovered, namely
\begin{align}
     \zeta(z) = z\,.\label{zetauv}
\end{align}
The behavior \eqref{zetauv} can be interpreted in dual theory as the restoration of conformal symmetry. 
\begin{figure}[htp!]%
    \centering
    {{\includegraphics[width=7.4cm]{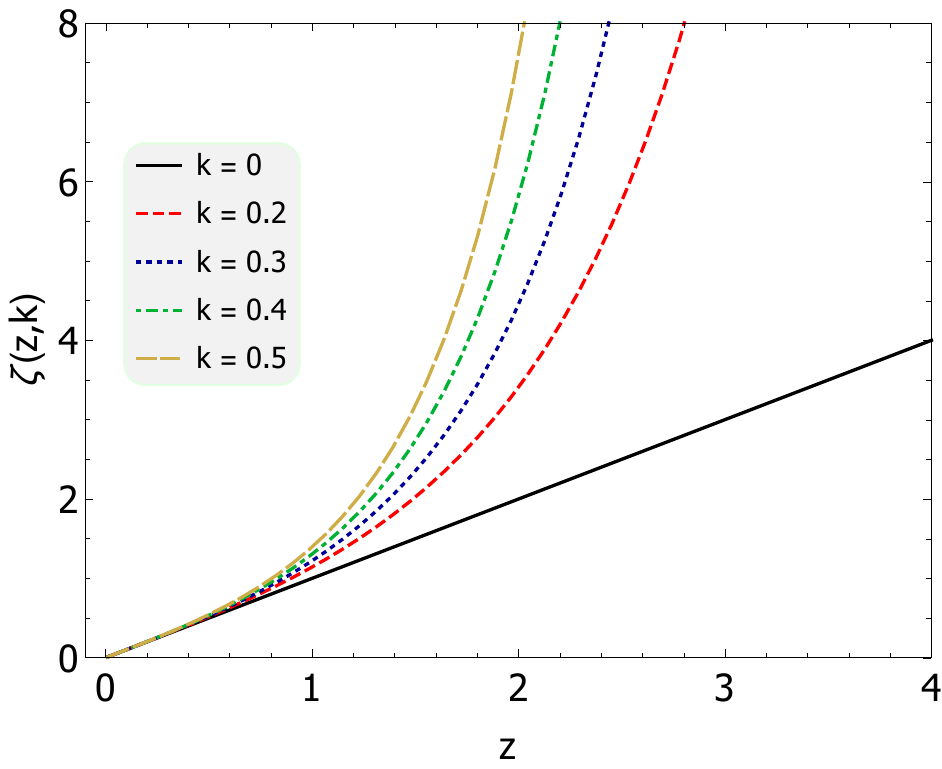}}}%
    \caption{Behavior of warp factor $\zeta(z)$ as function of holographic coordinate $z$ for several $k$ values.}%
    \label{Plot:zetamodelii}%
\end{figure}
By comparing the behavior of the warp factor for different values of $k$ in Fig.~\ref{Plot:zetamodelii} with the behavior of the dilaton for varying $\alpha$ (with $k=1$ fixed) in Fig.~\ref{Plot:DilatonWarpfactor}, we observe a competition in the breaking of conformal symmetry. While $k$ induces significant effects for $z \gtrsim 1$, the influence of $\alpha$ becomes more pronounced at larger values of $z$, as evidenced by the comparison between these figures.

The scale factor~\eqref{modelii} was originally employed to describe confinement in pure Einstein-dilaton gravity~\cite{Gursoy:2007er}, and we adopt it here in order to facilitate a direct comparison with existing results in the literature. In particular, our construction must reproduce the previous results in the limit $\alpha \to 0$. Substituting Eq.~\eqref{modelii} into \eqref{eqstarobinsky4}, we obtain the following modified dilaton:
\begin{align}
    \Phi(z,\alpha) =&\ \dfrac{1}{3}\int_{0}^{z}\, d\bar{z} \left[(10624 k^{3} \bar{z}^{4}\alpha + 2560 k^{4} \bar{z}^{6}\alpha)\exp\left(\frac{4}{3}k\,\bar{z}^{2}\right) + 81 k\left(1 + 8 \alpha\exp\left(\frac{4}{3}k\,\bar{z}^{2}\right)\right) \right.  \nonumber  \\
    &\left. + 12 k^{2} \bar{z}^{2}\left(3 + 968 \alpha \exp\left(\frac{4}{3}k\,\bar{z}^{2}\right)\right)\right]^{1/2}\,. \label{dilatonstarobinsky}
\end{align}
For $\alpha = 0$ (i.e., Einstein-dilaton gravity), the dilaton field in \eqref{dilatonstarobinsky} admits an analytical solution given by
\begin{align}
    \Phi(z) = \dfrac{1}{2}\sqrt{k}z\sqrt{9 + 4k z^{2}} + \dfrac{9}{4}\sinh^{-1}\left(\dfrac{2}{3}\sqrt{k}z\right)\,.\label{dilaton}
\end{align}
In both \eqref{dilatonstarobinsky} and \eqref{dilaton}, we obtain $\Phi (z) = 0$  when $k = 0$. The profile given in \eqref{dilaton} was also employed in Ref.~\cite{Ballon-Bayona:2024yuz}. Moreover, since the dilaton field is dimensionless and $k$ carries units of $\rm{GeV}^{2}$, the dimensionality of the free parameter $\alpha$ can be inferred from Eq.~\eqref{dilatonstarobinsky}. In particular, $\alpha$ must have units of $\rm{GeV}^{-2}$.

Since the differential equation~\eqref{eqstarobinsky4} exhibits a nontrivial dependence on both $\alpha$ and the radial coordinate, its solution must be obtained numerically. Consequently, our results will be expressed as functions of $z$ and $\alpha$. The left plot of Fig.~\ref{Plot:DilatonWarpfactor} displays the dilaton field for representative values of the Starobinsky parameter $\alpha$.

\begin{figure}[htp!]%
    \centering
    {{\includegraphics[width=7.4cm]{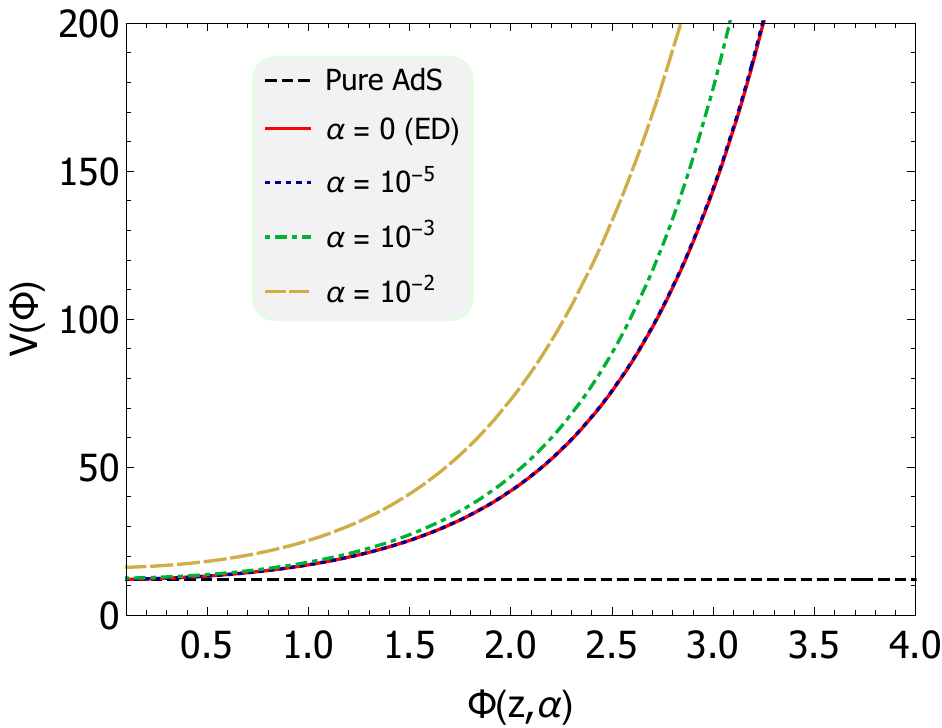}}}%
    \caption{Potential for Starobinsky-dilaton gravity.}%
    \label{Plot:VdilatonPotential}%
\end{figure}

By expressing \eqref{eqstarobinsky1} in terms of \eqref{modelii}, we obtain the potential of the model as follows:
\begin{align}
    V =& \bigg[\left(12 \zeta^{\prime\,\,2} - 3\zeta^{\prime\prime}\zeta\right)\left(1 + 2\alpha\left(8\zeta^{\prime\prime}\zeta - 20\zeta^{\prime\,2}\right)\right) + 2\alpha\left(8\zeta^{\prime\,\prime\,\prime\,\prime}\zeta^{\prime\,\prime\,\prime} - 32\zeta^{\prime\,\prime\, 2}\zeta^{2} - 24\zeta^{\prime\,\prime\,\prime}\zeta^{\prime}\zeta^{2}\right) \nonumber \\
    &- 12\alpha\zeta^{\prime}\zeta\left(8 \zeta^{\prime\,\prime\,\prime}\zeta - 32\zeta^{\prime}\zeta^{\prime\,\prime}\right)\bigg] + \alpha\left(8\zeta^{\prime\prime}\zeta - 20\zeta^{\prime\,2}\right)^{2}\,. \label{Starobinskypotential}
\end{align}
The asymptotic expansion for the potential \eqref{Starobinskypotential} is given by
\begin{align}
    V = 12 + 400\alpha + 12(3 + 280\alpha)z^{2} + \dfrac{160}{9}(3 + 623\alpha)z^{4} + \mathcal{O}(z^{5})\,,
\end{align}
such that when $\alpha = 0$, we recover the corresponding potential for the Einstein-dilaton model, that is,
\begin{align}
    V = 12 + 36z^{2} + \dfrac{160}{3}z^{4} + \mathcal{O}(z^{5})\,.
\end{align}
Figure \ref{Plot:VdilatonPotential} shows the behavior of the Starobinsky-dilaton gravity potential for several $\alpha$ values and the pure AdS case, in which the potential is $V(\Phi) = 12$ (represented by the black dashed line). Note that the Starobinsky-dilaton gravity effects begin to emerge from $\alpha \approx 10^{-3}$ (green dotdashed line). The potential increases due to $\alpha$-corrections.

\subsection{Confinement}
One of the main goals of this work is to investigate how the $\alpha R^2$ term influences the confinement criterion introduced in Ref.~\cite{Gursoy:2007er}. The confinement is characterized by the well-known area law of a rectangular Wilson loop, whose associated quark-antiquark potential energy, in the large-distance regime, is given by
\begin{equation}
    E(d) =\lambda \mathcal{F}(z^{\star})d\,,
\end{equation}
where $\lambda$ is the fundamental tension of the string, $\mathcal{F} =\exp(2A_{s})$ is the scalar factor of the model written in the string frame and $E(d)$ is the potential energy of the quark-antiquark pair as a function of the separation distance $d$ of the pair. For the fundamental tension of the string to be non-zero, the scale factor $\mathcal{F}$ must have a minimum located at $z = z^{\star}$. In our model, the function $\mathcal{F}$ is related with the warp factor $A_{s}$ which satisfies the equation
\begin{equation}
    A_{s}(z,\alpha) = A(z) + \dfrac{2}{3}\Phi(z,\alpha)\,.
\end{equation}
It is worth emphasizing that the $\alpha$-dependence of the warp factor arises through the dilaton field. The scale factor in the string frame, as a function of the holographic coordinate for the Starobinsky model (for different values of $\alpha$), is displayed in the right panel of Fig.~\ref{Plot:DilatonWarpfactor}. According to Ref.~\cite{Gursoy:2007er}, the confinement is equivalent to the existence of a minimum in the scale factor $\mathcal{F}$ described above, and as can be seen in Fig.~\ref{Plot:DilatonWarpfactor}, the corrections $\alpha R^{2}$ in our model maintain this important property.

\section{Vector mesons and Starobinsky-dilaton gravity}\label{section3}

\subsection{The action and field equations}
To study vector mesons, we will start with the following action written on the string frame
\begin{align}
    S = - \int d^{5}x\,\dfrac{1}{4g_{5}^{2}}\sqrt{-g_{s}}\,e^{-\Phi}F_{mn}^{(V)}F^{mn}_{(V)}\,, \label{actionvec}
\end{align}
where $\Phi$ is the dilaton field and $F_{mn}^{(V)}$ are the vector abelian field strengths given by
\begin{align}
    F_{mn}^{(V)} =\partial_{m}V_{n} - \partial_{n}V_{m}\,,
\end{align}
with $V_{m} =\tau^{a}V^{a}_{m}$ being the 5D vector gauge field with isospin symmetry $SU(2)$ where $\tau^{a}$ are the generators with $a = 1, 2, 3$. We are disregarding the axial sector of the field strengths because we want to describe only vector mesons. According to the holographic dictionary, gauge fields $A_{m}^{a}(x,z)$ are dual to currents $J_{\mu}^{a}(x)$ with conformal dimension $\Delta = 3$. In our case the vector gauge field $V_{m}(x,z)$ is dual to the vector current, $J_{\mu}^{V}(x)$, that is responsible for the creation of the vector meson. In addition, we are neglecting the non-abelian term of the field strengths, since we are interested in describing only the vector meson spectrum\footnote{However, the non-abelian term is important, for example, to study form factors \cite{Grigoryan:2007my, Grigoryan:2007vg}.}.  The dimensionless gauge coupling is given by $g_{5}^{2} = 12\pi^{2}/N_{c}$ to be compatible with the two-point function of the perturbative QCD \cite{Erlich:2005qh}. In our model, the vector gauge field also depends on the extra parameter $\alpha$, i.e., $V_{m} = V_{m}(x,z,\alpha)$. The 5D metric $g^{s}_{mn}$ in the string frame can be written as
\begin{equation}
    g^{s}_{mn} = e^{2A_{s}}\eta_{mn}\,. \label{metricstring}
\end{equation}
Plugging \eqref{metricstring} into \eqref{actionvec}, we can hence write the previous action as
\begin{align}
    S = - \int d^{5}x\, \dfrac{1}{4g_{5}^{2}}\,e^{A_{s} - \Phi}F_{mn}^{(V)}F^{mn}_{(V)}\,.
\end{align}

By varying the above action, one can obtain the following equation of motion
\begin{align}
    \partial_{m}\left(e^{A_{s} - \Phi}F^{mn}_{(V)}\right) = 0 .
\end{align}
Developing the last equation in components $\partial_{m} = (\partial_{z}, \partial_{\mu})$, $V_{m} = (V_{z}, V_{\mu})$, using the axial gauge $V_{z} = 0$ and performing a Lorentz decomposition in the vector gauge field $V_{m}(x, z)$, we get the equation of motion for the transverse vector gauge field
\begin{align}
    \left(\partial_{z} + A_{s}^{\prime} - \Phi^{\prime}\right)\partial_{z}V^{\mu}_{\perp} + \Box V^{\mu}_{\perp} = 0\,. \label{eqvectransverse}
\end{align}
Doing a Fourier transform
\begin{equation}
    V^{\mu} (x,z,\alpha) = \int d^{4}q\, e^{iq\cdot x} V^{\mu} (q,z)\,,
\end{equation}
equation \eqref{eqvectransverse} becomes
\begin{equation}
    \left(\partial_{z} + A_{s}^{\prime} - \Phi^{\prime}\right)\partial_{z}V^{\mu} - q^{2} V^{\mu} = 0\,.
\end{equation}

For simplicity from now on we will consider $V^{\mu}_{\perp} = V^{\mu}$. We remark that all quantities implicitly depend on the new parameter $\alpha$, but for brevity we will omit it in our written equations. The transverse vector gauge field has normalizable modes that satisfy the Sturm-Liouville equation
\begin{equation}
    \left[\partial_{z}\left(e^{A_{s} - \Phi}\partial_{z}\right) + m_{v_{n}}^{2}e^{A_{s} - \Phi}\right]v_{n}(z) = 0\,,\label{SturmLiouvillemodes}
\end{equation}
and they obey the orthonormality condition
\begin{equation}
    \int dz\, e^{A_{s} - \Phi}  v_{m}(z)v_{n}(z) =\delta_{mn}\,.
\end{equation}
In the next section we use this last result to obtain the spectrum of the vector mesons and their eigenfunctions.

\subsection{Vector meson spectrum}
In order to investigate the vector meson spectrum, we will perform the well-known Bogoliubov transformation on the normalizable modes
\begin{equation}
    v_{n} = e^{- B_{V}}\Psi_{n}^{(V)}\quad \text{with}\quad B_{V} =\dfrac{1}{2}(A_{s} - \Phi)\,.\label{Bogoliubov}
\end{equation}
Substituting \eqref{Bogoliubov} into \eqref{SturmLiouvillemodes}, we obtain the following Schrödinger equation
\begin{equation}
  -  \partial_{z}^{2}\Psi_{n}^{(V)}  + V_{V}\Psi_{n}^{(V)} =  m_{v^{n}}^{2}\Psi_{n}^{(V)}\,,
\end{equation}
where $\Psi_{n}^{(V)}$ are the eigenfunctions that describe the vector mesons, $m_{v^{n}}^{2}$ are the eigenvalues related with the spectrum and $V_{V}$ is the potential given by
\begin{equation}
    V_{V} = \partial_{z}^{2}B_{V} + (\partial_{z}B_{V})^{2}\,.\label{vecpotential}
\end{equation}

\begin{figure}[htp!]%
    \centering
    {{\includegraphics[width=7.4cm]{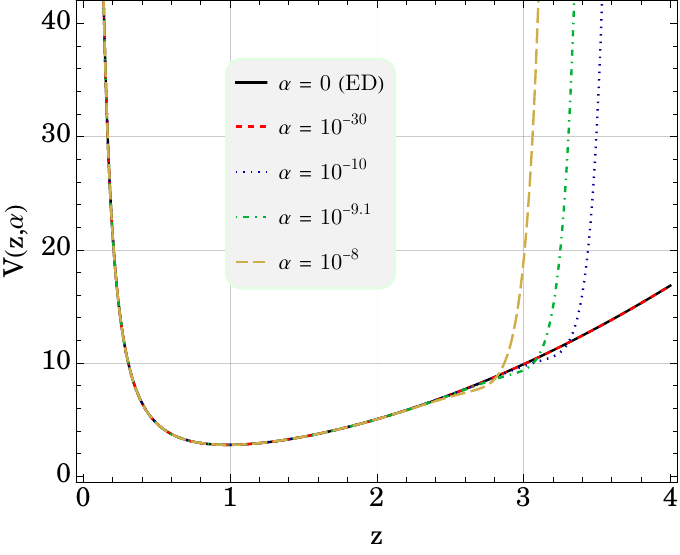}}}%
    \caption{Schrödinger potential as a function of the radial coordinate $z$ for usual Einstein-dilaton gravity ($\alpha = 0$) and for $R$-squared gravity as in Fig.~\ref{Plot:DilatonWarpfactor}.}%
    \label{Plot:Potential}%
\end{figure}

Figure \ref{Plot:Potential} displays the Schrödinger potential of vector mesons in the Starobinsky-dilaton gravity for some values of $\alpha$. One observes that, after a certain coordinate $z$, the potential decreases slightly and then suddenly grows as $\alpha$ increases.

In Table \ref{Table:VectorMesonMasses} we present our theoretical predictions for the vector mesons spectrum using $R$-squared gravity for four values of $\alpha$, and we compare with the experimental data as well as with the soft-wall model \cite{Karch:2006pv}. We adopt the following states for the vector mesons: $\rho(770)$, $\rho(1450)$ and $\rho(1700)$, following the particle data group (PDG) \cite{Workman:2022ynf}. In addition to these states, we also describe the state $\rho(1280)$ as reported by OBELIX Collaboration \cite{OBELIX:1997zla}. Although not a standard reference, this excitation plays an important role in the $\bar{p}p \rightarrow 2\pi^{+} 2\pi^{-}$ annihilation process and enables a more direct comparison with our theoretical predictions. It is worth mentioning that such resonance has been used in other holographic QCD models, see for instance Refs.~\cite{Gherghetta:2009ac, Bartz:2014oba, Ballon-Bayona:2023zal}.  All of these states have spin $1$, negative parity, and negative charge conjugation, i.e., $J^{PC}=1^{--}$. To describe the radial excitations of vector mesons in Starobinsky-dilaton gravity, we fix our infrared (IR) mass scale as $k = (0.348\, \text{GeV})^{2}$, resulting in the ground state rho meson $m_{\rho^{0}} = 0.776$ GeV for $\alpha = 0$ (i.e., pure Einstein gravity). Once the infrared input is determined for $\alpha =0$, we compute the vector meson spectrum for the aforementioned  $\alpha$ values to describe the effect of the $\alpha R^2$ term on the three excited states $n = 1, 2, 3$. Such a term has a little effect on the first excited state regardless of the considered $\alpha$ values. Nevertheless, the most significant impact of the parameter $\alpha$ on the mass occurs for the second and third excited states. Notice that for $\alpha = 10^{-8}$, our results for $m_{\rho^2}$ and $m_{\rho^3}$ are in good agreement with the experimental data. Note that the soft-wall model generally provides a better description of the vector meson spectrum than the Einstein-dilaton case ($\alpha = 0$). However, once Starobinsky-dilaton corrections are included, the agreement with experimental data improves significantly. In particular, for $\alpha = 10^{-8}$, the predicted spectrum becomes comparable and even more better than soft-wall model, in some cases. This improvement is especially evident for the higher excited states, notably for $m_{\rho^{3}}$, where the Starobinsky-dilaton result is closer to the experimental value than the corresponding soft-wall prediction.

\begin{table}[ht]
\centering
\begin{tabular}{l |c|c|c|c|c|c}
\hline 
\hline
$m_{\rho^{n}}$ & $\alpha = 0$ (ED) & $\alpha = 10^{-10}$ & $\alpha = 10^{- 9.1}$  & $\alpha = 10^{- 8}$ &  Soft-wall \cite{Karch:2006pv}  & Experimental \\
\hline 
 $m_{\rho^0}$ & 0.776  & 0.776  & 0.776  & 0.777 & 0.776 & $0.776 \pm 0.001$  \\
 $m_{\rho^1}$ &  1.041 & 1.043   &  1.047 & 1.058 & 1.097 & $1.282 \pm 0.037 $  \\
 $m_{\rho^2}$ &  1.258 &  1.276  & 1.296 & 1.336  & 1.344 & $1.465 \pm 0.025$ \\
 $m_{\rho^3}$ &  1.467 & 1.520 & 1.563  &  1.635  & 1.552 & $1.720 \pm 0.020$  \\
\hline\hline
\end{tabular}
\caption{Vector meson masses expressed in GeV for the ground state and three radial excitations $n = 1, 2,3$. The experimental data $m_{\rho^1}$ was extracted from Ref.~\cite{OBELIX:1997zla}. We investigated the vector meson spectrum considering Einstein-dilaton gravity $(\alpha = 0)$ and $\alpha = 10^{-10}, 10^{-9.1}, 10^{-8}$ for $f(R) = R + \alpha R^{2}$ gravity and compared against experimental data for the following states: $m_{\rho^{0}}$,  $m_{\rho^{2}}$ and $m_{\rho^{3}}$ \cite{Workman:2022ynf}.}
\label{Table:VectorMesonMasses}
\end{table}

The squared vector meson masses as a function of the radial excitation number are displayed in the left panel of Fig.~\ref{Plot:SquaMassesWaveFun} for the chosen values of $\alpha$.  For $\alpha = 0$, corresponding to Einstein-dilaton gravity, the spectrum exhibits an approximately linear dependence on the radial excitation number. The main consequence of the quadratic contribution $\alpha R^2$ on the spectrum is to increases the squared masses as $\alpha$ increases, and it becomes the spectrum nonlinear with the radial excitation number. Particularly, the more considerable impact takes place for the most highly excited states when  $\alpha = 10^{-8}$. Other models, such as $D4/D8$ \cite{BallonBayona:2009ar} and $D3/D7$ \cite{Kruczenski:2003be}, describe the square of the rho meson masses as nonlinear with radial excitation. This behavior is also observed in simpler models, such as the hard-wall model \cite{Boschi-Filho:2002xih, Boschi-Filho:2002wdj}. \\

\medskip

{\bf Asymptotic analysis of the effective potential}

\medskip 

To obtain a deeper understanding of which term dominates the Schrödinger potential for vector mesons and contributes to the spectrum's nonlinearity, we will carry out an asymptotic analysis of the potential. The potential of vector mesons \eqref{vecpotential} depends on derivatives of $B_{V}$ given by \eqref{Bogoliubov}. The asymptotic behavior of $B_{V}$ is
\begin{align}
    B_{V}(z,\alpha) = -\dfrac{1}{2}\log z - \dfrac{1}{2}\sqrt{1 + 8\alpha} z - \dfrac{2}{3}z^{2} + \dfrac{(-3 - 1040\alpha)z^{3}}{81\sqrt{1 + 8\alpha}} + \cdots \label{Bvasymptotic}
\end{align}
Consequently, the potential is given by
\begin{align}
    V_{V}(z, \alpha) =&\ \dfrac{3}{4z^{2}} + \dfrac{\sqrt{1 + 8\alpha}}{2z} + \dfrac{1}{4}(1 + 8\alpha) + \dfrac{(15 - 896\alpha)z}{27\sqrt{1 + 8\alpha}} + \dfrac{5}{27}(3 + 208\alpha)z^{2} + \nonumber  \\
    &+ \dfrac{(45 - 41040\,\alpha + 701056\,\alpha^{2})z^{3}}{243(1 + 8\alpha)^{3/2}} + \dfrac{2224\,\alpha\,z^{4}}{27} + \cdots\label{VVasymptotic}
\end{align}
In the UV regime, for small $z$, the leading term in the potential \eqref{VVasymptotic} is
\begin{align}
    V_{V}(z,\alpha) = \dfrac{3}{4z^{2}}\,,\label{VVasymptoticUV}
\end{align}
indicating that there is no Starobinsky dilaton gravity correction in this regime. 

The higher-order terms in $z$ appearing in the asymptotic expansion \eqref{VVasymptotic} are responsible for the nonlinear behavior of the vector meson spectrum for the three values of $\alpha \neq0$ presented in Table \ref{Table:VectorMesonMasses}. The asymptotic expansion allowed us to predict which term of the potential is relevant to the Starobinsky-dilaton gravity. Nonetheless, to understand how this affects the spectrum, we will use the WKB approximation following Ref.~\cite{Gursoy:2007er}, namely
\begin{align}\label{npiEq}
    n \pi = \int_{z_{1}}^{z_{2}}\sqrt{m^{2} - V_{V}(z,\alpha)}\,dz\,,
\end{align}
where $z_{1}$ and $z_{2}$ are the turning points. As we have seen, the effects are appreciable in the potential only for large $z$. The dominant term of the full Schrödinger potential in the large-$z$ regime is
\begin{align}\label{largzasymptoticpotential}
    V_{V}(z\to \infty, \alpha)\sim \alpha\,\exp\left({\frac{4}{3}z^{2}}\right) .
\end{align}
When $m^{2}$ is large enough, the turning points are $z_{1} = 0$ and $z_{2} \sim \sqrt{\frac{3}{4}\ln\left(\frac{m^{2}}{\alpha}\right)}$ so that
\begin{align}\label{wkb}
    n \pi = \int_{0}^{\sqrt{\frac{3}{4}\ln\left(\frac{m^{2}}{\alpha}\right)}}\sqrt{ m^{2} - \alpha\,\exp\left({\frac{4}{3}z^{2}}\right) }\, dz\,,
\end{align}
which leads us to the following relationship between mass and radial excitation number:
\begin{align}
    m_{n}^{2}\sim \dfrac{n^{2}}{\ln\left(\dfrac{n^{2}}{\alpha}\right)}. \label{mvsnrelation}
\end{align}
Note that the result \eqref{mvsnrelation} expresses a non-linear relation between mass and radial excitation number and has a logarithmic dependence on $\alpha$. It is possible to recover linear behavior if we expand the potential \eqref{largzasymptoticpotential} into a power series, truncate the series to the quadratic order and plugging into the WKB approximation \eqref{npiEq}, which would correspond to the small-$n$ regime, where the corrections in $\alpha$ are not significant. According to the result \eqref{mvsnrelation}, the corrections $\alpha R^{2}$ significantly modify the vector meson spectrum for large $n$, which can be observed in the left plot of Fig.~\ref{Plot:SquaMassesWaveFun}.

\subsection{Wave functions and decay constants}
\begin{figure}[htp!]%
    \centering
    {{\includegraphics[width=7.4cm]{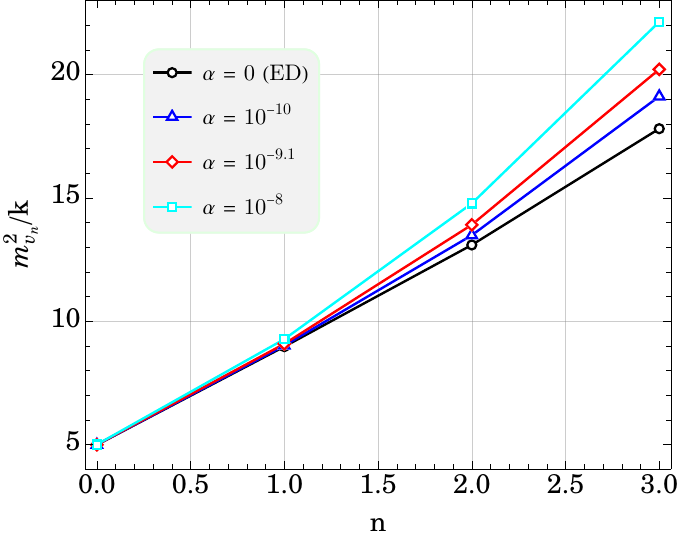}}
    {\includegraphics[width=7.49cm]{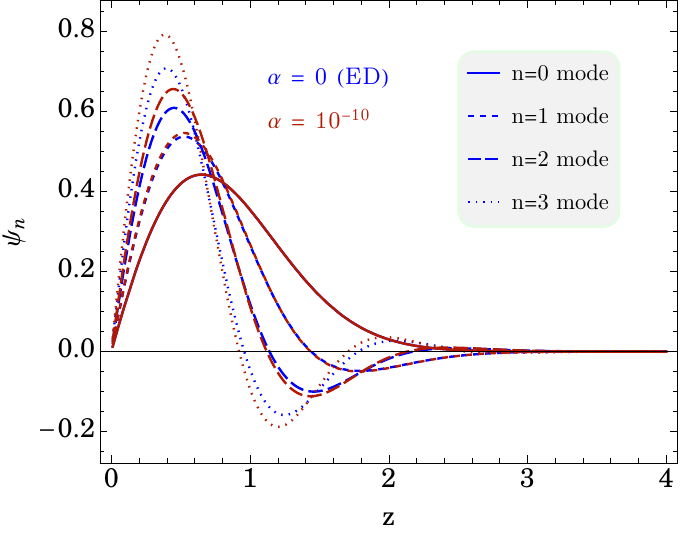}}}%
    \caption{Left panel: Squared masses for vector mesons as a function of the radial excitation number in Starobinsky-dilaton gravity for four values of $\alpha$. Right plot: Wave functions of vector mesons, where blue and red curves stand for the pure Einstein-dilaton case and the Starobinsky-dilaton model, respectively.}%
    \label{Plot:SquaMassesWaveFun}%
\end{figure}

In the previous section we described the potential and vector meson spectrum. To solve the Schrödinger equation and obtain the eigenvalues numerically, we require that the eigenfunctions vanish at large $z$, i.e., far from the boundary
\begin{equation}
    \lim_{z\to \infty} \sqrt{z}\Psi_{n}^{(V)} = 0\,.
\end{equation}
The radial behavior of the wave functions describing vector mesons is displayed in the right panel of Fig.~\ref{Plot:SquaMassesWaveFun} for pure Einstein-dilaton gravity (blue curves) and Starobinsky-dilaton gravity (red lines). The differences between the wave functions in these two models become more pronounced in the higher excited states, where the $\alpha R^2$ corrections are more noticeable, leading to larger wave peaks compared to those of the pure Einstein-dilaton model. This trend aligns with the effects on the potential and spectrum discussed in previous sections, where the Starobinsky term $\alpha R^{2}$ increasingly affects the higher excited states.

The decay constants of vector mesons are related with the following matrix element
\begin{align}
    \bra{0}J^{a}_{\mu}(0)\ket{v_{n}^{b}} =\delta^{ab}f_{v^n}\epsilon_{\mu}\,,
\end{align}
where $f_{n}$ are the decay constants and $\epsilon_{\mu}$ the polarization vector. These decay constants describe decay from a vector meson state to the hadronic vacuum, and they can be calculated from the two-point function obtained from Sturm-Liouvile theory. Following Ref.~\cite{Ballon-Bayona:2024yuz}, such constants can be written as
\begin{align}
    f_{\rho^n} =\dfrac{1}{g_{5}}\left[e^{A_{s} - \Phi}\partial_{z}v_{n}\right]_{z =\epsilon}\,.
\end{align}
In Table \ref{Table:DecayvectorMesons}, we present our results for the decay constants of vector mesons in the context of Starobinsky-dilaton gravity against experimental data and Einstein-dilaton gravity ($\alpha = 0$). We consider corrections for decay constants of vector mesons for  $\alpha = 10^{-10}$. Similar to the spectrum, the impact of $\alpha$-corrections tends to increase the decay constants for the more excited states in comparison to the Einstein-dilaton gravity.

\begin{table}[ht]
\centering
\begin{tabular}{|c|c|c|c|c|c}
\hline 
\hline
$f_{\rho^{n}}^{1/2}$ & $\alpha = 0$ (ED) & $\alpha = 10^{-10}$ & Experimental \\
\hline 
$f_{\rho^0}^{1/2}$ &  0.2193  & 0.2193 & $0.3462 \pm 0.0014$  \\
$f_{\rho^1}^{1/2}$ & 0.2646 & 0.2670  & $0.433 \pm 0.013$  \\
$f_{\rho^2}^{1/2}$ & 0.3012 & 0.3144  &  \\
$f_{\rho^3}^{1/2}$ & 0.3443 &  0.3694 &  \\
\hline\hline
\end{tabular}
\caption{Decay constants of vector meson expressed in GeV for the ground state and first radial excitations $n = 1,2,3$. We investigated the vector meson spectrum considering Einstein-dilaton gravity $(\alpha = 0)$ and $\alpha = 10^{-10}$ for $f(R) = R + \alpha R^{2}$ gravity and compared against experimental results \cite{Donoghue:1992dd}.}
\label{Table:DecayvectorMesons}
\end{table}

\section{Conclusions}\label{section4}

In this work we have described the $R$-squared gravity effects on the vector meson spectrum. We showed that the Starobinsky model influences in a non-trivial way such spectrum and reproduces very satisfactorily the experimental data for the most excited states. To describe the impact of the Starobinsky term on the vector meson sector, we first generalized the Einstein-dilaton model by introducing the arbitrary function $f(R)$ in the action and deriving the field equations in the metric formalism. Accordingly, for the specific function \eqref{starobinsky}, we had the Starobinsky-dilaton gravity model in $\rm AdS_{5}$ spacetime. We then expressed the field equation in terms of the inverse scale factor $\zeta(z)$ and of dilaton field $\Phi(z)$. We have also shown and analyzed the $\alpha$-corrections for the dilaton potential in \eqref{Starobinskypotential}, depicted in Fig.~\ref{Plot:VdilatonPotential}. Since the equation \eqref{eqstarobinsky4} is of first order in the dilaton and fourth order in the inverse of the scale factor, we choose to use the inverse of the scale factor as presented in \eqref{modelii} inspired by previous works \cite{Ballon-Bayona:2017sxa, Ballon-Bayona:2024yuz}. This scale factor profile satisfies the confinement criteria discussed in \cite{Gursoy:2007er}. Using $\zeta(z)$ we obtained the dilaton field \eqref{dilatonstarobinsky} which depends explicitly on the parameter $\alpha$ and solved the integral numerically.
\

After obtaining the dilaton field and warp factor with $\alpha R^{2}$ corrections, we showed that our model preserves the property of the scale factor, $\mathcal{F} =\exp(2A_{s})$, having a minimum. Then, by deriving the Schrödinger equation for vector mesons we obtained the potential, as shown in Fig. \ref{Plot:Potential}, as well as the wave function and the spectrum presented in Fig. \ref{Plot:SquaMassesWaveFun}, both including the higher-curvature $\alpha R^{2}$ contribution. To obtain the vector meson spectrum, we have fixed the IR mass scale as $k = (0.348\, \text{GeV})^{2}$ for $\alpha = 0$. This is the input to our model in order to reproduce the ground state mass of the rho vector meson. Once fixed, we presented the ground state and three radial excitation masses of rho meson in Table \ref{Table:VectorMesonMasses} for $\alpha = 0, 10^{-10}, 10^{-9.1}$ and $10^{-8}$. The effects are weakest for ground state ($m_{\rho^{0}}$) and for the first excited state ($m_{\rho^{1}}$), however, the effect becomes stronger for second ($m_{\rho^{2}}$) and third excited state ($m_{\rho^{3}}$). When we compare the our results with the experimental data, we observe that $m_{\rho^{2}} = 1.336$ and $m_{\rho^{3}} = 1.635$ for $\alpha = 10^{-8}$  are close to $1.465 \pm 0.025$ and $1.720 \pm 0.020$ respectively. In Table \ref{Table:DecayvectorMesons}, we present our model's prediction for the decay constants for vector mesons at $\alpha = 10^{-10}$ compared with experimental data and the Einstein-dilaton model. Similar to the spectrum, the effects of the Starobinsky model on the decay constants are most significant for excited states. Therefore, based on $f(R)$ gravity, we have developed a modified Einstein-dilaton model with two parameters: $\alpha$, which governs the Starobinsky gravity effects, and $k$, which introduces a mass scale into the model and show that it is possible to construct a model with $\alpha R^{2}$ effects in the rho-meson spectrum, preserving the confinement criteria compatible with experimental data.

Although the inclusion of $\alpha R^{2}$ corrections in our $f(R)$-dilaton model yields a nonlinear $\rho$-meson spectrum consistent with experimental data, the present analysis is limited to the $\rho$ sector alone. A comprehensive extension of the framework to other vector mesons, as well as to scalar, pseudoscalar, and axial-vector states, would require a more detailed treatment incorporating, for instance, chiral and flavor symmetry breaking effects. Such considerations lie beyond the scope of the current work and constitute promising directions for future investigations, some of which are briefly outlined below.

Overall, this work highlights the relevance of extending the Einstein-dilaton framework and explores its implications for the vector meson spectrum. In future studies, we plan to apply the present $f(R)$-dilaton gravity model to the spectra of the other meson channels, as well as to baryons. It would also be worthwhile to consider alternative choices of $f(R)$, such as the Hu-Sawicki model \cite{Hu2007} and power-law $f(R)$ models \cite{Capozziello2006, Capozziello2007, Pretel2022CQG}, in order to further investigate their impact on mesons and baryons.

\section*{Acknowledgments}

The authors gratefully acknowledge the valuable suggestions and comments of Alfonso Ballon-Bayona and Henrique Boschi-Filho. A.S.S.~Jr is supported by the Conselho Nacional de Desenvolvimento Cientifico e Tecnologico (CNPq). JMZP acknowledges support from ``Fundação Carlos Chagas Filho de Amparo à Pesquisa do Estado do Rio de Janeiro'' -- FAPERJ, Process SEI-260003/000308/2024.


\bibliographystyle{utphys}

\bibliography{Starobinskyhqcd}

@article{Maldacena:1997re,
    author = "Maldacena, Juan Martin",
    title = "{The Large N limit of superconformal field theories and supergravity}",
    eprint = "hep-th/9711200",
    archivePrefix = "arXiv",
    reportNumber = "HUTP-97-A097, HUTP-98-A097",
    doi = "10.4310/ATMP.1998.v2.n2.a1",
    journal = "Adv. Theor. Math. Phys.",
    volume = "2",
    pages = "231--252",
    year = "1998"
}

@article{Witten:1998qj,
    author = "Witten, Edward",
    title = "{Anti-de Sitter space and holography}",
    eprint = "hep-th/9802150",
    archivePrefix = "arXiv",
    reportNumber = "IASSNS-HEP-98-15",
    doi = "10.4310/ATMP.1998.v2.n2.a2",
    journal = "Adv. Theor. Math. Phys.",
    volume = "2",
    pages = "253--291",
    year = "1998"
}

@article{Gubser:1998bc,
    author = "Gubser, S. S. and Klebanov, Igor R. and Polyakov, Alexander M.",
    title = "{Gauge theory correlators from noncritical string theory}",
    eprint = "hep-th/9802109",
    archivePrefix = "arXiv",
    reportNumber = "PUPT-1767",
    doi = "10.1016/S0370-2693(98)00377-3",
    journal = "Phys. Lett. B",
    volume = "428",
    pages = "105--114",
    year = "1998"
}

@book{Ammon:2015wua,
    author = "Ammon, Martin and Erdmenger, Johanna",
    title = "{Gauge/gravity duality}: {Foundations and applications}",
    doi = "10.1017/CBO9780511846373",
    isbn = "978-1-107-01034-5, 978-1-316-23594-2",
    publisher = "Cambridge University Press",
    address = "Cambridge",
    month = "4",
    year = "2015"
}

@article{Karch:2006pv,
    author = "Karch, Andreas and Katz, Emanuel and Son, Dam T. and Stephanov, Mikhail A.",
    title = "{Linear confinement and AdS/QCD}",
    eprint = "hep-ph/0602229",
    archivePrefix = "arXiv",
    reportNumber = "BUHEP-06-02, INT-PUB-06-04",
    doi = "10.1103/PhysRevD.74.015005",
    journal = "Phys. Rev. D",
    volume = "74",
    pages = "015005",
    year = "2006"
}

@article{Grigoryan:2007vg,
    author = "Grigoryan, Hovhannes R. and Radyushkin, Anatoly V.",
    title = "{Form Factors and Wave Functions of Vector Mesons in Holographic QCD}",
    eprint = "hep-ph/0703069",
    archivePrefix = "arXiv",
    reportNumber = "JLAB-THY-07-624",
    doi = "10.1016/j.physletb.2007.05.044",
    journal = "Phys. Lett. B",
    volume = "650",
    pages = "421--427",
    year = "2007"
}

@article{Grigoryan:2007my,
    author = "Grigoryan, H. R. and Radyushkin, A. V.",
    title = "{Structure of vector mesons in holographic model with linear confinement}",
    eprint = "0706.1543",
    archivePrefix = "arXiv",
    primaryClass = "hep-ph",
    reportNumber = "JLAB-THY-07-652",
    doi = "10.1103/PhysRevD.76.095007",
    journal = "Phys. Rev. D",
    volume = "76",
    pages = "095007",
    year = "2007"
}

@article{Grigoryan:2007wn,
    author = "Grigoryan, H. R. and Radyushkin, A. V.",
    title = "{Pion form-factor in chiral limit of hard-wall AdS/QCD model}",
    eprint = "0709.0500",
    archivePrefix = "arXiv",
    primaryClass = "hep-ph",
    reportNumber = "JLAB-THY-07-708",
    doi = "10.1103/PhysRevD.76.115007",
    journal = "Phys. Rev. D",
    volume = "76",
    pages = "115007",
    year = "2007"
}

@article{Abidin:2008ku,
    author = "Abidin, Zainul and Carlson, Carl E.",
    title = "{Gravitational form factors of vector mesons in an AdS/QCD model}",
    eprint = "0801.3839",
    archivePrefix = "arXiv",
    primaryClass = "hep-ph",
    doi = "10.1103/PhysRevD.77.095007",
    journal = "Phys. Rev. D",
    volume = "77",
    pages = "095007",
    year = "2008"
}

@article{Abidin:2008hn,
    author = "Abidin, Zainul and Carlson, Carl E.",
    title = "{Gravitational Form Factors in the Axial Sector from an AdS/QCD Model}",
    eprint = "0804.0214",
    archivePrefix = "arXiv",
    primaryClass = "hep-ph",
    doi = "10.1103/PhysRevD.77.115021",
    journal = "Phys. Rev. D",
    volume = "77",
    pages = "115021",
    year = "2008"
}

@article{Abidin:2009aj,
    author = "Abidin, Zainul and Carlson, Carl E.",
    title = "{Strange hadrons and kaon-to-pion transition form factors from holography}",
    eprint = "0908.2452",
    archivePrefix = "arXiv",
    primaryClass = "hep-ph",
    doi = "10.1103/PhysRevD.80.115010",
    journal = "Phys. Rev. D",
    volume = "80",
    pages = "115010",
    year = "2009"
}

@article{DaRold:2005mxj,
    author = "Da Rold, Leandro and Pomarol, Alex",
    title = "{Chiral symmetry breaking from five dimensional spaces}",
    eprint = "hep-ph/0501218",
    archivePrefix = "arXiv",
    reportNumber = "UAB-FT-578",
    doi = "10.1016/j.nuclphysb.2005.05.009",
    journal = "Nucl. Phys. B",
    volume = "721",
    pages = "79--97",
    year = "2005"
}

@article{Hung:2011xb,
    author = "Hung, Ling-Yan and Myers, Robert C. and Smolkin, Michael",
    title = "{On Holographic Entanglement Entropy and Higher Curvature Gravity}",
    eprint = "1101.5813",
    archivePrefix = "arXiv",
    primaryClass = "hep-th",
    doi = "10.1007/JHEP04(2011)025",
    journal = "JHEP",
    volume = "04",
    pages = "025",
    year = "2011"
}

@article{Brigante:2007nu,
    author = "Brigante, Mauro and Liu, Hong and Myers, Robert C. and Shenker, Stephen and Yaida, Sho",
    title = "{Viscosity Bound Violation in Higher Derivative Gravity}",
    eprint = "0712.0805",
    archivePrefix = "arXiv",
    primaryClass = "hep-th",
    reportNumber = "CAS-KITPC-ITP-025, MIT-CTP-3918, SU-ITP-07-22",
    doi = "10.1103/PhysRevD.77.126006",
    journal = "Phys. Rev. D",
    volume = "77",
    pages = "126006",
    year = "2008"
}

@article{Cho:2002hq,
    author = "Cho, Y. M. and Neupane, Ishwaree P.",
    title = "{Anti-de Sitter black holes, thermal phase transition and holography in higher curvature gravity}",
    eprint = "hep-th/0202140",
    archivePrefix = "arXiv",
    reportNumber = "SNUTP-2002-04",
    doi = "10.1103/PhysRevD.66.024044",
    journal = "Phys. Rev. D",
    volume = "66",
    pages = "024044",
    year = "2002"
}

@article{Santos:2021orr,
    author = "Santos, Fabiano F. and Capossoli, Eduardo Folco and Boschi-Filho, Henrique",
    title = "{AdS/BCFT correspondence and BTZ black hole thermodynamics within Horndeski gravity}",
    eprint = "2105.03802",
    archivePrefix = "arXiv",
    primaryClass = "hep-th",
    doi = "10.1103/PhysRevD.104.066014",
    journal = "Phys. Rev. D",
    volume = "104",
    number = "6",
    pages = "066014",
    year = "2021"
}

@article{Sotiriou2010,
  title = {$f({R})$ theories of gravity},
  author = {T.~P.~Sotiriou and V.~Faraoni},
  journal = {Rev. Mod. Phys.},
  volume = {82},
  issue = {1},
  pages = {451--497},
  numpages = {0},
  year = {2010},
  doi = {10.1103/RevModPhys.82.451}
}

@article{DeFelice2010,
  title = {$f({R})$ Theories},
  author = {A.~De Felice and S.~Tsujikawa},
  journal = {Living Rev. Relativ.},
  volume = {13},
  pages = {3},
  year = {2010},
  doi = {10.12942/lrr-2010-3}
}

@article{Capozziello2011,
  title = "Extended Theories of Gravity",
  journal = "Phys. Rep.",
  volume = "509",
  number = "4",
  pages = "167 - 321",
  year = "2011",
  doi = "https://doi.org/10.1016/j.physrep.2011.09.003",
  author = "Salvatore Capozziello and Mariafelicia {De Laurentis}"
}

@article{Nojiri2011,
  title = {Unified cosmic history in modified gravity: From $F({R})$ theory to Lorentz non-invariant models},
  journal = {Phys. Rep.},
  volume = {505},
  number = {2},
  pages = {59-144},
  year = {2011},
  doi = {https://doi.org/10.1016/j.physrep.2011.04.001},
  author = {Shin’ichi Nojiri and Sergei D. Odintsov}
}

@article{Clifton2012,
  title = "Modified gravity and cosmology",
  journal = "Phys. Rep.",
  volume = "513",
  number = "1",
  pages = "1-189",
  year = "2012",
  doi = "https://doi.org/10.1016/j.physrep.2012.01.001",
  author = "Timothy Clifton and others"
}

@article{Nojiri2017,
  title = {Modified gravity theories on a nutshell: Inflation, bounce and late-time evolution},
  journal = {Phys. Rep.},
  volume = {692},
  pages = {1-104},
  year = {2017},
  doi = {https://doi.org/10.1016/j.physrep.2017.06.001},
  author = {S. Nojiri and S.D. Odintsov and V.K. Oikonomou}
}

@article{NOJIRI2020,
  title = {Unifying inflation with early and late-time dark energy in $F({R})$ gravity},
  journal = {Phys. Dark Univ.},
  volume = {29},
  pages = {100602},
  year = {2020},
  doi = {https://doi.org/10.1016/j.dark.2020.100602},
  author = {Shin’ichi Nojiri and Sergei D. Odintsov and V.K. Oikonomou}
}

@article{Carroll2004,
  title = {Is cosmic speed-up due to new gravitational physics?},
  author = {Carroll, Sean M. and Duvvuri, Vikram and Trodden, Mark and Turner, Michael S.},
  journal = {Phys. Rev. D},
  volume = {70},
  issue = {4},
  pages = {043528},
  numpages = {5},
  year = {2004},
  doi = {10.1103/PhysRevD.70.043528}
}

@article{Song2007,
  title = {Large scale structure of $f({R})$ gravity},
  author = {Song, Yong-Seon and Hu, Wayne and Sawicki, Ignacy},
  journal = {Phys. Rev. D},
  volume = {75},
  issue = {4},
  pages = {044004},
  numpages = {10},
  year = {2007},
  doi = {10.1103/PhysRevD.75.044004},
}

@article{Ali2010,
  title = {Background cosmological dynamics in $f({R})$ gravity and observational constraints},
  author = {Ali, Amna and Gannouji, Radouane and Sami, M. and Sen, Anjan A.},
  journal = {Phys. Rev. D},
  volume = {81},
  issue = {10},
  pages = {104029},
  numpages = {6},
  year = {2010},
  doi = {10.1103/PhysRevD.81.104029}
}

@article{Ribeiro2024,
    author = {B. Ribeiro and A. Bernui and M. Campista},
    title = {Cosmological constraints on the 
${R}^2$-corrected {A}ppleby–{B}attye model},
    journal = {Eur. Phys. J. C},
    volume = {84},
    pages = {114},
    year = {2024},
    doi = {10.1140/epjc/s10052-024-12437-x}
}

@article{Starobinsky1980,
  title = "A new type of isotropic cosmological models without singularity",
  journal = "Phys. Lett. B",
  volume = "91",
  number = "1",
  pages = "99 - 102",
  year = "1980",
  issn = "0370-2693",
  doi = "10.1016/0370-2693(80)90670-X",
  author = "A. A. Starobinsky"
}

@article{Akrami2020,
	author = {Akrami, Y. and others},
	title = {Planck 2018 results - {X}. Constraints on inflation},
	DOI= "10.1051/0004-6361/201833887",
	journal = {A\&A},
	year = 2020,
	volume = 641,
	pages = "A10",
}

@article{Astashenok2017,
  doi = {10.1088/1361-6382/aa8971},
  year = {2017},
  volume = {34},
  number = {20},
  pages = {205008},
  author = {Artyom V Astashenok and Sergei D Odintsov and Álvaro de la Cruz-Dombriz},
  title = {The realistic models of relativistic stars in $f({R}) = {R} + \alpha {R}^2$ gravity},
  journal = {Class. Quantum Grav.}
}

@article{Olmo2020,
  title = {Stellar structure models in modified theories of gravity: Lessons and challenges},
  journal = {Phys. Rep.},
  volume = {876},
  pages = {1-75},
  year = {2020},
  doi = {https://doi.org/10.1016/j.physrep.2020.07.001},
  author = {Gonzalo J. Olmo and Diego Rubiera-Garcia and Aneta Wojnar}
}

@article{Astashenok2020,
  title = {Extended gravity description for the {GW190814} supermassive neutron star},
  journal = {Phys. Lett. B},
  volume = {811},
  pages = {135910},
  year = {2020},
  doi = {https://doi.org/10.1016/j.physletb.2020.135910},
  author = {A.V. Astashenok and others}
}

@article{Astashenok2021,
  title = {Causal limit of neutron star maximum mass in $f({R})$ gravity in view of {GW}190814},
  journal = {Phys. Lett. B},
  volume = {816},
  pages = {136222},
  year = {2021},
  doi = {https://doi.org/10.1016/j.physletb.2021.136222},
  author = {A.V. Astashenok and others}
}

@article{Nobleson2022,
  title = {Comparison of perturbative and non-perturbative methods in $f({R})$ gravity},
  author = {K. Nobleson and A. Ali and S. Banik},
  journal = {Eur. Phys. J. C},
  volume = {82},
  pages = {32},
  year = {2022},
  doi = {https://doi.org/10.1140/epjc/s10052-021-09969-x}
}

@article{Jimenez2022,
  doi = {10.1088/1475-7516/2022/07/017},
  year = {2022},
  volume = {2022},
  number = {07},
  pages = {017},
  author = {José C. Jiménez and others},
  title = {${R}^2$-gravity quark stars from perturbative {QCD}},
  journal = {JCAP}
}

@article{Pretel2022JCAP,
	doi = {10.1088/1475-7516/2022/09/058},
	year = 2022,
	volume = {2022},
	number = {09},
	pages = {058},
	author = {J. M. Z. Pretel and others},
	title = {Charged quark stars in metric $f({R})$ gravity},
	journal = {JCAP}
}

@article{Kehagias2015,
  title = {Black hole solutions in ${R}^2$ gravity},
  author = {A. Kehagias and C. Kounnas and D. Lüst and A. Riotto},
  journal = {JHEP},
  volume = {2015},
  number = {05},
  pages = {143},
  year = {2015},
  doi = {https://doi.org/10.1007/JHEP05(2015)143}
}

@article{Cognola2015,
  title = {Thermodynamics of topological black holes in ${R}^{2}$ gravity},
  author = {Cognola, Guido and Rinaldi, Massimiliano and Vanzo, Luciano and Zerbini, Sergio},
  journal = {Phys. Rev. D},
  volume = {91},
  issue = {10},
  pages = {104004},
  numpages = {9},
  year = {2015},
  doi = {10.1103/PhysRevD.91.104004},
}

@article{Li2024,
  title = {Holographic images of an {A}d{S} black hole within the framework of $f({R})$ gravity theory},
  author = {Guo-Ping Li and Ke-Jian He and Xin-Yun Hu and Qing-Quan Jiang},
  journal = {Front. Phys.},
  volume = {19},
  pages = {54202},
  year = {2024},
  doi = {https://doi.org/10.1007/s11467-024-1393-8}
}

@article{MartinContreras:2020cyg,
    author = "Martin Contreras, Miguel Angel and Vega, Alfredo",
    title = "{Nonlinear Regge trajectories with AdS/QCD}",
    eprint = "2004.10286",
    archivePrefix = "arXiv",
    primaryClass = "hep-ph",
    doi = "10.1103/PhysRevD.102.046007",
    journal = "Phys. Rev. D",
    volume = "102",
    number = "4",
    pages = "046007",
    year = "2020"
}

@article{Braga:2025lah,
    author = "Braga, Nelson R. F. and Ferreira, Yan F.",
    title = "{Non linear Regge trajectories of quarkonia from holography}",
    eprint = "2512.20371",
    archivePrefix = "arXiv",
    primaryClass = "hep-ph",
    month = "12",
    year = "2025"
}

@article{Afonin:2009xi,
    author = "Afonin, S. S.",
    title = "{AdS/QCD models describing a finite number of excited mesons with Regge spectrum}",
    eprint = "0903.0322",
    archivePrefix = "arXiv",
    primaryClass = "hep-ph",
    doi = "10.1016/j.physletb.2009.03.073",
    journal = "Phys. Lett. B",
    volume = "675",
    pages = "54--58",
    year = "2009"
}

@article{Chen:2022flh,
    author = "Chen, Jiao-Kai",
    title = "{Revisiting the pion Regge trajectories}",
    eprint = "2203.02981",
    archivePrefix = "arXiv",
    primaryClass = "hep-ph",
    doi = "10.1016/j.nuclphysb.2022.115911",
    journal = "Nucl. Phys. B",
    volume = "983",
    pages = "115911",
    year = "2022"
}

@article{Ballon-Bayona:2017sxa,
    author = "Ballon-Bayona, Alfonso and Boschi-Filho, Henrique and Mamani, Luis A. H. and Miranda, Alex S. and Zanchin, Vilson T.",
    title = "{Effective holographic models for QCD: glueball spectrum and trace anomaly}",
    doi = "10.1103/PhysRevD.97.046001",
    journal = "Phys. Rev. D",
    volume = "97",
    number = "4",
    pages = "046001",
    year = "2018"
}

@article{Gursoy:2007cb,
    author = "Gursoy, U. and Kiritsis, E.",
    title = "{Exploring improved holographic theories for QCD: Part I}",
    eprint = "0707.1324",
    archivePrefix = "arXiv",
    primaryClass = "hep-th",
    reportNumber = "CPHT-RR027-0507",
    doi = "10.1088/1126-6708/2008/02/032",
    journal = "JHEP",
    volume = "02",
    pages = "032",
    year = "2008"
}

@article{Gursoy:2007er,
    author = "Gursoy, U. and Kiritsis, E. and Nitti, F.",
    title = "{Exploring improved holographic theories for QCD: Part II}",
    eprint = "0707.1349",
    archivePrefix = "arXiv",
    primaryClass = "hep-th",
    reportNumber = "CPHT-RR028-0507",
    doi = "10.1088/1126-6708/2008/02/019",
    journal = "JHEP",
    volume = "02",
    pages = "019",
    year = "2008"
}

@article{Gursoy:2008za,
    author = "Gursoy, U. and Kiritsis, E. and Mazzanti, L. and Nitti, F.",
    title = "{Holography and Thermodynamics of 5D Dilaton-gravity}",
    eprint = "0812.0792",
    archivePrefix = "arXiv",
    primaryClass = "hep-th",
    reportNumber = "CPHT-RR088-1108, SPIN-08-57, ITP-UU-08-74",
    doi = "10.1088/1126-6708/2009/05/033",
    journal = "JHEP",
    volume = "05",
    pages = "033",
    year = "2009"
}

@article{Ballon-Bayona:2023zal,
    author = "Ballon-Bayona, Alfonso and Frederico, Tobias and Mamani, Luis A. H. and de Paula, Wayne",
    title = "{Dynamical holographic QCD model for spontaneous chiral symmetry breaking and confinement}",
    eprint = "2308.07503",
    archivePrefix = "arXiv",
    primaryClass = "hep-ph",
    doi = "10.1103/PhysRevD.108.106016",
    journal = "Phys. Rev. D",
    volume = "108",
    number = "10",
    pages = "106016",
    year = "2023"
}

@article{Ballon-Bayona:2021tzw,
    author = "Ballon-Bayona, Alfonso and Mamani, Luis A. H. and Miranda, Alex S. and Zanchin, Vilson T.",
    title = "{Effective holographic models for QCD: Thermodynamics and viscosity coefficients}",
    eprint = "2103.14188",
    archivePrefix = "arXiv",
    primaryClass = "hep-th",
    doi = "10.1103/PhysRevD.104.046013",
    journal = "Phys. Rev. D",
    volume = "104",
    number = "4",
    pages = "046013",
    year = "2021"
}

@article{Gursoy:2009kk,
    author = "Gursoy, Umut and Kiritsis, Elias and Michalogiorgakis, Georgios and Nitti, Francesco",
    title = "{Thermal Transport and Drag Force in Improved Holographic QCD}",
    eprint = "0906.1890",
    archivePrefix = "arXiv",
    primaryClass = "hep-ph",
    reportNumber = "CPHT-RR052-0609, ITP-UU-09-22, SPIN-09-21",
    doi = "10.1088/1126-6708/2009/12/056",
    journal = "JHEP",
    volume = "12",
    pages = "056",
    year = "2009"
}

@article{Erlich:2005qh,
    author = "Erlich, Joshua and Katz, Emanuel and Son, Dam T. and Stephanov, Mikhail A.",
    title = "{QCD and a holographic model of hadrons}",
    doi = "10.1103/PhysRevLett.95.261602",
    journal = "Phys. Rev. Lett.",
    volume = "95",
    pages = "261602",
    year = "2005"
}

@article{Workman:2022ynf,
    author = "Workman, R. L. and others",
    collaboration = "Particle Data Group",
    title = "{Review of Particle Physics}",
    doi = "10.1093/ptep/ptac097",
    journal = "PTEP",
    volume = "2022",
    pages = "083C01",
    year = "2022"
}

@book{Donoghue:1992dd,
    author = "Donoghue, J. F. and Golowich, E. and Holstein, Barry R.",
    title = "{Dynamics of the standard model}",
    doi = "10.1017/CBO9780511524370",
    publisher = "CUP",
    volume = "2",
    year = "2014"
}

@article{BallonBayona:2009ar,
    author = "Ballon Bayona, C. A. and Boschi-Filho, Henrique and Braga, Nelson R. F. and Torres, Marcus A. C.",
    title = "{Form factors of vector and axial-vector mesons in holographic D4-D8 model}",
    eprint = "0911.0023",
    archivePrefix = "arXiv",
    primaryClass = "hep-th",
    doi = "10.1007/JHEP01(2010)052",
    journal = "JHEP",
    volume = "01",
    pages = "052",
    year = "2010"
}

@article{Kruczenski:2003be,
    author = "Kruczenski, Martin and Mateos, David and Myers, Robert C. and Winters, David J.",
    title = "{Meson spectroscopy in AdS / CFT with flavor}",
    eprint = "hep-th/0304032",
    archivePrefix = "arXiv",
    doi = "10.1088/1126-6708/2003/07/049",
    journal = "JHEP",
    volume = "07",
    pages = "049",
    year = "2003"
}

@article{Boschi-Filho:2002xih,
    author = "Boschi-Filho, Henrique and Braga, Nelson R. F.",
    title = "{Gauge / string duality and scalar glueball mass ratios}",
    eprint = "hep-th/0212207",
    archivePrefix = "arXiv",
    doi = "10.1088/1126-6708/2003/05/009",
    journal = "JHEP",
    volume = "05",
    pages = "009",
    year = "2003"
}

@article{Boschi-Filho:2002wdj,
    author = "Boschi-Filho, Henrique and Braga, Nelson R. F.",
    title = "{QCD / string holographic mapping and glueball mass spectrum}",
    eprint = "hep-th/0209080",
    archivePrefix = "arXiv",
    doi = "10.1140/epjc/s2003-01526-4",
    journal = "Eur. Phys. J. C",
    volume = "32",
    pages = "529--533",
    year = "2004"
}

@article{Ballon-Bayona:2024yuz,
    author = "Ballon-Bayona, Alfonso and Junior, Ad{\~a}o S. da Silva",
    title = "{Nucleons and vector mesons in a confining holographic QCD model}",
    eprint = "2402.17950",
    archivePrefix = "arXiv",
    primaryClass = "hep-ph",
    doi = "10.1103/PhysRevD.109.094050",
    journal = "Phys. Rev. D",
    volume = "109",
    number = "9",
    pages = "094050",
    year = "2024"
}

@article{OBELIX:1997zla,
    author = "Bertin, A. and others",
    collaboration = "OBELIX",
    title = "{Study of anti-p p --\ensuremath{>} 2pi+ 2pi- annihilation from S states}",
    doi = "10.1016/S0370-2693(97)01189-1",
    journal = "Phys. Lett. B",
    volume = "414",
    pages = "220--228",
    year = "1997"
}

@article{Hu2007,
  title = {Models of $f({R})$ cosmic acceleration that evade solar system tests},
  author = {Hu, Wayne and Sawicki, Ignacy},
  journal = {Phys. Rev. D},
  volume = {76},
  issue = {6},
  pages = {064004},
  numpages = {13},
  year = {2007},
  doi = {10.1103/PhysRevD.76.064004}
}

@article{Capozziello2006,
  doi = {10.1088/1475-7516/2006/08/001},
  year = {2006},
  volume = {2006},
  number = {08},
  pages = {001},
  author = {S Capozziello and V F Cardone and A Troisi},
  title = {Dark energy and dark matter as curvature effects?},
  journal = {JCAP}
}

@article{Capozziello2007,
  doi = {10.1088/0264-9381/24/8/013},
  year = {2007},
  volume = {24},
  number = {8},
  pages = {2153},
  author = {S Capozziello and A Stabile and A Troisi},
  title = {Spherically symmetric solutions in $f({R})$ gravity via the Noether symmetry approach},
  journal = {Class. Quantum Grav.}
}

@article{Pretel2022CQG,
  doi = {10.1088/1361-6382/ac7a88},
  year = {2022},
  volume = {39},
  number = {15},
  pages = {155003},
  author = {Juan M Z Pretel and Sérgio B Duarte},
  title = {Anisotropic quark stars in $f({R}) ={R}^{1+\epsilon}$ gravity},
  journal = {Class. Quantum Grav.}
}

@article{Gherghetta:2009ac,
    author = "Gherghetta, Tony and Kapusta, Joseph I. and Kelley, Thomas M.",
    title = "{Chiral symmetry breaking in the soft-wall AdS/QCD model}",
    eprint = "0902.1998",
    archivePrefix = "arXiv",
    primaryClass = "hep-ph",
    doi = "10.1103/PhysRevD.79.076003",
    journal = "Phys. Rev. D",
    volume = "79",
    pages = "076003",
    year = "2009"
}

@article{Bartz:2014oba,
    author = "Bartz, Sean P. and Kapusta, Joseph I.",
    title = "{Dynamical three-field AdS/QCD model}",
    eprint = "1406.3859",
    archivePrefix = "arXiv",
    primaryClass = "hep-ph",
    doi = "10.1103/PhysRevD.90.074034",
    journal = "Phys. Rev. D",
    volume = "90",
    number = "7",
    pages = "074034",
    year = "2014"
}

\end{document}